\documentclass[12pt]{iopart} 
\usepackage{iopams}
\usepackage{graphicx}
\usepackage{textcomp}
\newcommand{\yps}{YbPt$_2$Sn}
\newcommand{\ypi}{YbPt$_2$In}
\begin{document}
\title[Unusual weak magnetic exchange in different structure types]{Unusual weak magnetic exchange in two different structure types: \yps\ and \ypi}
\author{T Gruner$^1$, D Jang$^1$, A Steppke$^1$, M Brando$^{1,2}$, F Ritter$^3$, C Krellner$^3$ and C Geibel$^1$}
\address{$^1$ Max Planck Institute for Chemical Physics of Solids, 01187 Dresden, Germany}
\address{$^2$ I. Physikalisches Institut, University of G\"ottingen, 37077 G\"ottingen, Germany}
\address{$^3$ Institute of Physics, Goethe University Frankfurt, 60438 Frankfurt/M, Germany}
\eads{Thomas.Gruner@cpfs.mpg.de}
\begin{abstract}
We present the structural, magnetic, thermodynamic, and transport properties of the two new compounds \yps\ and \ypi. X-ray powder diffraction shows that they crystallize in different structure types, the hexagonal ZrPt$_2$Al and the cubic Heusler type, respectively. Despite quite different lattice types, both compounds present very similar magnetic properties: a stable trivalent Yb$^{3+}$, no evidence for a sizeable Kondo interaction, and very weak exchange interactions with a strength below 1\,K as deduced from specific heat $C(T)$. Broad anomalies in $C(T)$ suggest short range magnetic ordering at about 250\,mK and 180\,mK for \yps\ and \ypi, respectively. The weak exchange and the low ordering temperature result in a large magnetocaloric effect as deduced from the magnetic field dependence of $C(T)$, making these compounds interesting candidates for magnetic cooling. In addition we found in \ypi\ evidences for a charge density wave transition at about 290\,K. The occurrence of such transitions within several RET$_2$X compound series (RE\,=\,rare earth, T\,=\,noble metal, X\,=\,In, Sn) is analyzed.
\end{abstract}
%
\maketitle
\section{Introduction}
In the last decade Yb-based intermetallic compounds have attracted considerable attention because of the observation of unconventional and unique properties in quite a number of cases. Prominent examples are: i)~the heavy fermion compound YbRh$_2$Si$_2$, which is one of the most interesting system for studying quantum critical points \cite{Trov00a,Cust03}; ii)~YbAlB$_4$, the only Yb-based unconventional superconductor, where superconductivity seems to emerge from valence fluctuations instead of the commonly assumed antiferromagnetic fluctuations \cite{Naka08,Mats11}, and iii)~the series YbT$_2$Zn$_{20}$, which present the highest mass renormalization observed up to now in strongly correlated electron systems \cite{Tori07,Onuk11}. These unusual properties emerge from the instability of the 4$f$ shell of Yb and the associated valence instability of this element. In the divalent Yb$^{2+}$ state the 4$f$ shell is fully occupied and therefore non-magnetic. In the trivalent Yb$^{3+}$ state one hole in the 4$f$ shell yield a local 4$f$ moment with $S = 1/2$, $L = 3$ and thus $J = 7/2$ according to Hund's rules. However, in quite a number of Yb-based compounds a strong hybridization between the 4$f$ and the conduction electrons results in an intermediate valent state or in the formation of a Kondo lattice. In such systems the interaction of the 4$f$ electrons with the conduction electrons results in strong electronic correlation effects and a huge renormalization of the quasi particles at the Fermi level, hence the name heavy fermion.\\
The discovery of these interesting phenomena motivates the search for new Yb-based intermetallic compounds with unusual properties. In the present paper we report the synthesis of the new compounds \yps\ and \ypi\ as well as the investigation of their structural and magnetic properties. While the related system YbPd$_2$Sn has been widely investigated quite some time ago and was found to show coexistence of classical superconductivity below $T_c = 2.3$\,K and antiferromagnetism of localized trivalent Yb below $T_N = 0.22$\,K \cite{Ishi82,Aoki00}, neither \yps\ nor \ypi\ have yet been reported. Our results revealed these two compounds to crystallize in very different structure types. Nevertheless, they exhibit similar magnetic properties: in both compounds Yb is in a stable trivalent state, without any evidence for the strong correlation effects searched for. In contrast, the exchange interactions are exceptionally weak, resulting in a quite large calculated magnetocaloric effect. Thus instead of being of interest for the field of strongly correlated electron systems, both compounds turned out to be promising candidates for magnetic cooling. Furthermore, in \ypi\ we observe at about 290\,K evidence for a structural phase transition, likely of charge density wave (CDW) type. Similar structural phase transitions have already been reported in several RET$_2$X compound series (RE\,=\,rare earth, T\,=\,noble metal, X\,=\,In, Sn) crystalizing in the cubic Heusler structure type. A comparative analysis of these series of compounds reveals a systematic trend: the occurrence of this phase transition seems to be connected with the change from the hexagonal ZrPt$_2$Al to the cubic Heusler structure type within a compound series. This change is governed by the size of the RE atom and by the main quantum number of the noble metal, while the total number of valence electrons does not seem to be important.
\section{Sample preparation}
The REPt$_2$Sn and REPt$_2$In series of compounds were first reported in 1987 by A.\,E.~Dwight for RE\,=~Gd, Tb, Er, Tm and Gd\,-\,Ho, respectively \cite{Dwig87}. All samples were found to crystallize in the hexagonal ZrPt$_2$Al structure type. For the Gd-based compounds this was later confirmed by K.~Zhang and L.~Chen \cite{Zhan92} as well as by B.~Heying et al. \cite{Heyi09}. However, we did not find any report on the Yb- or on the Lu-based homologues. Because of the huge difference in the melting points of the pure elements ($T_{\rm Pt}\approx2040$\,K, $T_{\rm Yb}\approx1100$\,K, $T_{\rm Sn}\approx500$\,K and $T_{\rm In}\approx430$\,K) we prepared the polycrystalline samples in a two-stage arc-melting process under ultrapure argon atmosphere. In a first stage an appropriate amount of Yb and of the low melting elements Sn or In was melted to a small button. In the second step Pt was added and melted with the prereacted YbSn/YbIn mixture. In this step the samples were repeatedly melted and turned over to enhance homogeneity. For both compounds the total weight loss after the whole procedure was about 4\,wt\%. This can safely be attributed to the evaporation of Yb due to its low boiling point of 1469\,K. Therefore, an excess of 10\% to 15\% Yb was used in the initial composition to compensate for this loss. No impurity phases were detected in as-cast \yps, except for minor islands of Yb$_{12}$Pt$_{48}$Sn$_{40}$ as indicated by energy-dispersive microprobe analysis. Therefore, the sample were investigated in the as-cast state. Since preliminary results indicated a lower phase purity in the as-cast \ypi\ ingots, they were put into a Tungsten boat, wrapped with Tantalum foil and annealed at 1073\,K for 100\,h under 800\,mbar purified argon. Microprobe analysis revealed that the annealed \ypi\ samples still include some unreacted Pt and small amounts of Yb-deficient phases. The precise control of the Yb content during the preparation of these compounds is likely the main problem in enhancing the sample quality. The homologue non-magnetic counterparts LuPt$_2$Sn and LuPt$_2$In have been synthesized by the same procedure.
\section{Experimental details}
Room temperature X-ray powder diffraction data were recorded on a STOE Stadip instrument in transmission mode and using Cu $K\alpha$1 radiation. X-ray powder spectra of freshly crushed \ypi\ showed broadened Bragg peaks (shown in figure~\ref{fig01}({\bf{b}})). Annealing the powder again resulted in a significant narrowing of the peaks, but at the same time the size and number of foreign phase peaks increase, likely due to surface oxidation (not shown). The low-temperature diffraction experiments were carried out using a Siemens D500 powder diffractometer, equipped with a closed cycle helium refrigerator. The accessible sample temperature range extends from 10\,K to 300\,K. The samples were placed directly on a temperature controlled Cu block by means of Apiezon N grease to ensure good thermal contact. For temperature control a silicon diode temperature sensor was used that was mounted inside this Cu block directly beneath the sample. At each temperature the powdered sample was carefully thermalized before starting the diffraction scan. The diffraction scans were recorded with Cu $K\alpha$1 radiation. The resistivity $\rho(T)$ and specific heat capacity $C(T)$ were measured in the $T$ range of 350\,mK to 400\,K using a commercial Quantum Design (QD) PPMS equipped with a $^3$He option. $C(T)$ in the millikelvin range down to around 50\,mK was determined in a $^3$He/$^4$He dilution refrigerator using a compensated heat pulse method. The magnetization and susceptibility measurements above 1.8\,K were carried out using a QD SQUID VSM, while low-temperature $M(H)$ and $\chi(T)$ data were taken using a QD MPMS equipped with a $^3$He option.
\section{Experimental results and discussion}
\subsection{Structure at room temperature}
%
%
\begin{figure}[tbh]
	\flushright
  \includegraphics[width=0.85\textwidth]{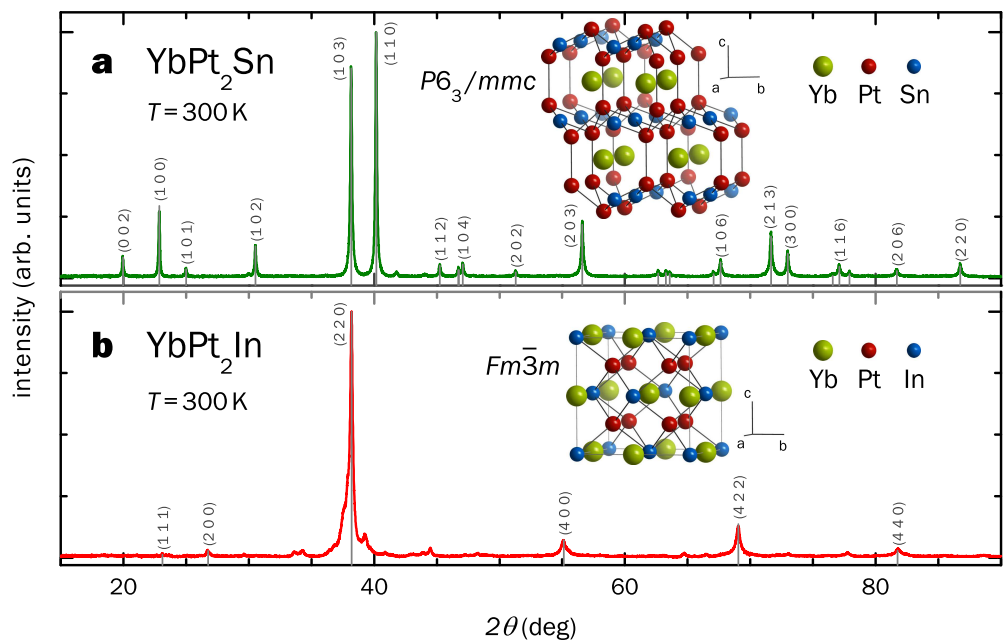}
  \caption{Powder X-ray diffraction patterns of \yps\ and \ypi\ at room temperature. The grey bars at the bottom of the diagrams indicate the location of Bragg peaks of the hexagonal ZrPt$_2$Al and of the cubic Heusler phase structure types, respectively. The pictures of the structures illustrate the quasi-2D character with alternating Yb and Pt$_2$Sn layers for \yps, contrasting the 3D-network in cubic \ypi. ({\bf{a}}) For \yps\ all reflexes except for two faint lines were indexed, the indices are only given for the strongest peaks. ({\bf{b}}) In \ypi\ the pattern is dominated by the strong peaks corresponding to the $a/2$ bcc underlying structure, but the (111) and (200) peaks corresponding to the ordering of Yb, Pt and In in the Heusler phase are well observed. The origin of the additional peaks is discussed in the text.
  }
 \label{fig01}
\end{figure}
%
Surprisingly, an examination of the X-ray powder diffraction patterns of \yps\ and \ypi\ shows that they crystallize in completely different structure types (figure~\ref{fig01}). In \yps\ all peaks, except for two very weak ones ($2\theta \approx 42$° and 44°) could be assigned to the hexagonal ZrPt$_2$Al structure type, space group {\it P}6{$_3$}{\it /mmc}, with the lattice parameters $a=4.4862(5)$\,\AA\ and $c=8.8881(8)$\,\AA. The identified peaks are marked with grey bars and the strongest ones labeled with the corresponding indices. The small number and weak intensities of unidentified peaks confirms an almost phase pure sample as already deduced from microprobe investigations. The same result was also observed for LuPt$_2$Sn, with only slightly smaller lattice parameters $a=4.4803(5)$\,\AA\ and $c=8.8765(9)$\,\AA. Comparing volume per formula unit within the REPt$_2$Sn compound series using in addition the data of \cite{Dwig87} reveals a continuous decrease with increasing atomic number of the rare earth: $V_{\rm {Er}}=78.30$\,\AA$^3$, $V_{\rm {Tm}}=77.94$\,\AA$^3$, $V_{\rm {Yb}}=77.46$\,\AA$^3$ and $V_{\rm {Lu}}=77.15$\,\AA$^3$. This indicates Yb to be in the trivalent state. The picture of the \yps\ structure shown in figure~\ref{fig01}({\bf{a}}) evidences some quasi-two-dimensional character with an alternation of Yb and Pt$_2$Sn layers.\\
In contrast, the few strong Bragg reflections in the X-ray pattern of \ypi\ (figure~\ref{fig01}({\bf{b}})) initially indicate a simple body-centered cubic (bcc) underlying structure. However, ordering of Yb, Pt and In in a cubic Heusler structure type ($L2_1$, {\it Fm}{\=3}{\it m}) is clearly evidenced by the appearance of the (111) and (200) peaks. The X-ray pattern of \ypi\ is less clean than that of \yps, in accordance with the larger amount of foreign phases observed with microprobe. However, as discussed later, some of the weak peaks initially suspected to belong to foreign phases turned out to be intrinsic superstructure peaks connected with the formation of a CDW. A similar results was also observed for LuPt$_2$In. The main difference is a slight decrease of the lattice parameter from $a=6.655(3)$\,\AA\ for the Yb-based to $a=6.6484(7)$\,\AA\ for the Lu-based compound. Thus, our results evidence that within the REPt$_2$In series of compounds the structure changes from the hexagonal ZrPt$_2$Al type to the cubic Heusler type somewhere in-between RE\,=\,Ho and RE\,=\,Yb. No data are yet available for RE\,=\,Er and Tm. The volume per formula unit of HoPt$_2$In, $V_{\rm {Ho}}=79.10$\,\AA$^3$, is significantly larger than that of \ypi\ and LuPt$_2$In, $V_{\rm {Yb}}=73.68$\,\AA$^3$ and $V_{\rm {Lu}}=73.47$\,\AA$^3$, respectively, the difference being much larger than expected from the lanthanide contraction. This indicates that the Heusler structure type is denser than the hexagonal one. On the other hand, the volume difference between \ypi\ and LuPt$_2$In is smaller than that for the Sn homologues. This indicates Yb to be in the trivalent state in \ypi, too. The three-dimensional network of cubic \ypi\ is shown in figure~\ref{fig01}({\bf{b}}).
\subsection{Weak exchange interactions}
%
%
\begin{figure}[tbh]
	\flushright
  \includegraphics[width=0.85\textwidth]{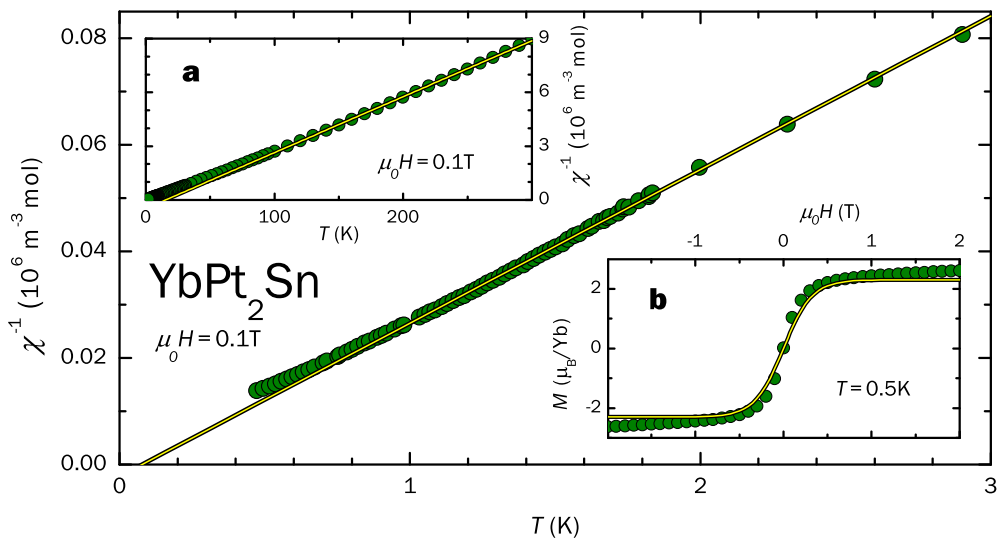}
  \caption{Temperature dependence of the inverse susceptibility $\chi^{-1}$ of \yps\ in a field of 0.1\,T at high (inset ({\bf{a}})) and low temperatures (central part). In both $T$-ranges $\chi(T)$ follows a Curie-Weiss law with effective moments ($\mu_{\rm eff}^{\rm high\,T} = 4.52\,\mu_{\rm B}$ and $\mu_{\rm eff}^{\rm low\,T} = 4.70\,\mu_{\rm B}$, respectively) close to the value for a stable trivalent Yb$^{3+}$ state (4.53\,$\mu_{\rm B}$). The Weiss temperatures for the high temperature fit ($\Theta^{\rm high\,T} = -15$\,K) is likely related to the crystal electric field (CEF) splitting. The low $T$ fit results in a very small negative value $\Theta^{\rm low\,T} = -0.08$\,K, giving a first hint toward the exchange being very weak and dominantly ferromagnetic. Inset ({\bf{b}}) shows the magnetization measured as a function of field at low $T = 0.5$\,K. The comparison with an effective $S = 1/2$ Brillouin function with $M_{\rm sat} = 2.3\,\mu_{\rm B} / {\rm Yb}$ is a further evidence for a very weak exchange interaction.
  }
 \label{fig02}
\end{figure}
\begin{figure}[tb]
	\flushright
  \includegraphics[width=0.85\textwidth]{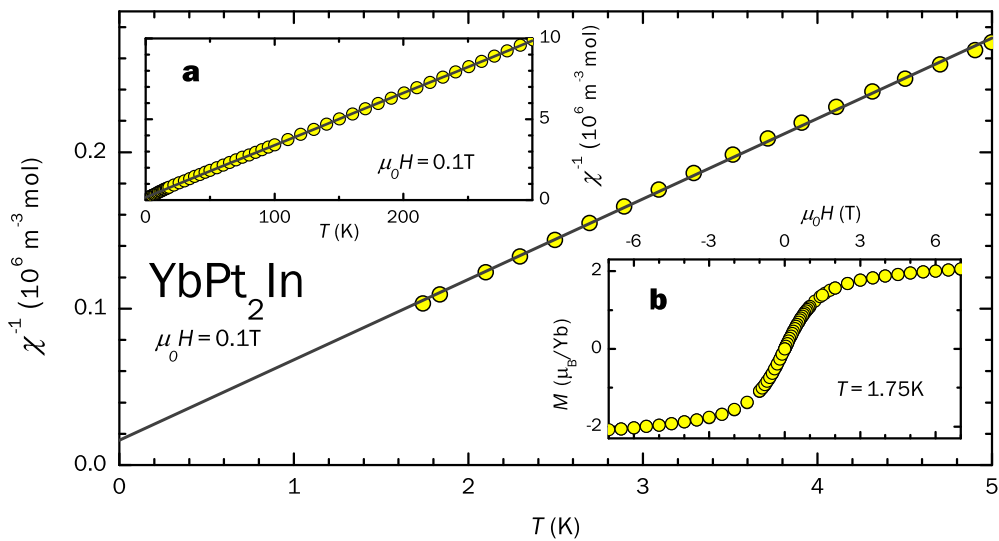}
  \caption{Temperature dependence of $\chi^{-1}$ of \ypi\ in a magnetic field of 0.1\,T. Analysis similar to that for \yps\ in figure~\ref{fig02} yield $\mu_{\rm eff}^{\rm high\,T} = 4.44\,\mu_{\rm B}$, $\mu_{\rm eff}^{\rm low\,T} = 3.52\,\mu_{\rm B}$, and $\Theta^{\rm low\,T} = 0.31$\,K. Inset ({\bf{b}}) shows the magnetization measured as a function of field at $T = 1.75$\,K.
  }
 \label{fig02a}
\end{figure}
\begin{figure}[tb]
	\flushright
  \includegraphics[width=0.85\textwidth]{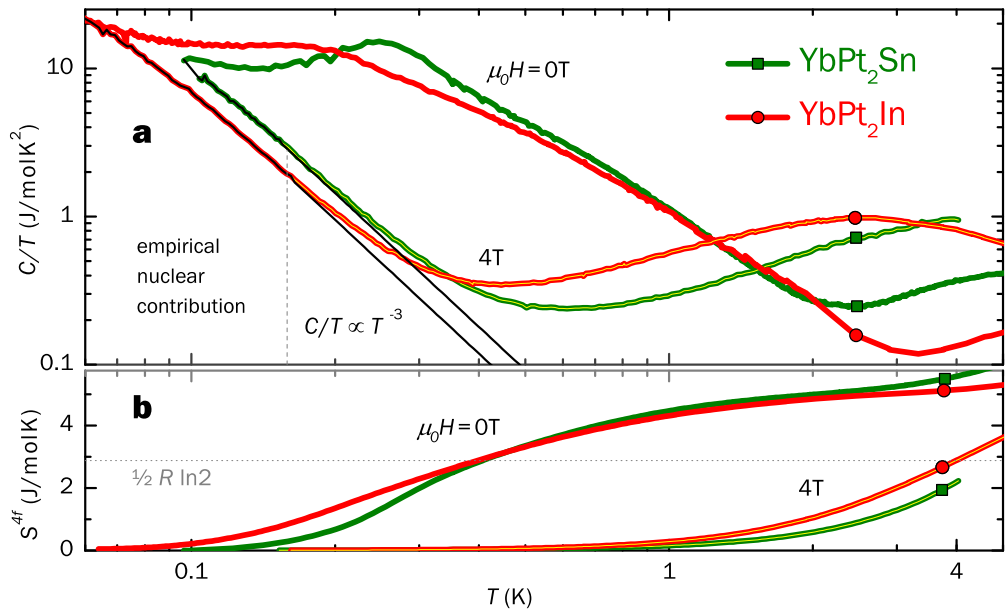}
  \caption{({\bf{a}}) Temperature dependence of the measured specific heat of \yps\ and \ypi\ in fields of 0\,T and 4\,T. At $\mu_{0}H = 0$\,T the strong increase below 2\,K and peaks near 240\,mK and 180\,mK for the Sn and the In compound, respectively, indicate magnetic order of local moments dominated by short range correlations. A magnetic field of 4\,T results in a Schottky anomaly at around 3\,K because of the Zeeman splitting of the 4$f$ states. The increase at low $T$ is due to the nuclear Yb contribution. ({\bf{b}}) Temperature dependence of the entropy deduced from $C(T)$ data after subtraction of a nuclear and a phonon contribution. Above 1\,K $S^{4f}(T)$ levels out at nearly $R$\,ln2 indicating a well separated crystal field ground state doublet for both systems. In both compounds $1/2\,R$\,ln2 is reached at $T = 0.4$\,K, indicating the overall strength of the exchange to be below 1\,K.
  }
 \label{fig03}
\end{figure}

%
The temperature dependence of the inverse susceptibility $\chi^{-1}(T)$ of \yps\ in an external field of $\mu_{0}H = 0.1$\,T is shown in inset~({\bf{a}}) of figure~\ref{fig02}. It displays an almost linear behavior, which can be nicely fitted above 100\,K to a Curie-Weiss ($\chi^{-1} = (T + \Theta) / C_{\rm CW}$) law with an effective moment ($\mu_{\rm eff}^{\rm high\,T} = 4.52\,\mu_{\rm B}$) very close to the value for the stable trivalent Yb$^{3+}$ state (4.53\,$\mu_{\rm B}$). In Yb-based systems the Weiss temperature (here $\Theta^{\rm high\,T} = -15$\,K) obtained from the fit at high temperatures is usually determined by crystal electric field (CEF) effects since exchange is typically very weak due to the small de Gennes factor. A more reliable estimation of exchange strength can be obtained from a Curie-Weiss fit of $\chi^{-1}$ in the low temperature range (shown in the central part of figure~\ref{fig02}). The fitted low $T$ Weiss temperature $\Theta^{\rm low\,T} = -0.08$\,K is a first indicator of weak Yb-Yb intersite exchange. The negative sign suggest a dominant ferromagnetic (FM) interaction, but considering the very small value and the large range of the extrapolation, this conclusion is not very reliable. Towards lowest temperatures $\chi^{-1}$ shows the beginning of an upturn, but this might be the result of the onset of saturation due to magnetic energy becoming comparable to thermal energy. This is confirmed by the field dependence of the magnetization at $T = 0.5$\,K shown in figure~\ref{fig02}({\bf{b}}). The experimental data, which show a well-defined saturation above 0.5\,T to a value slightly larger than $2\,\mu_{\rm B} / {\rm Yb}$, are compared with an effective $S = 1/2$ Brillouin function for $T = 0.5$\,K with the only free parameter, the saturation magnetization, fixed to $M_{\rm sat} = 2.3\,\mu_{\rm B} / {\rm Yb}$. The nice agreement with this single ion model confirms the weakness of the exchange interactions. The saturation moment is of the size typically expected for the doublets resulting from the splitting of the $J = 7/2$ multiplet of Yb$^{3+}$ induced by the CEF.\\
The susceptibility and magnetization data of \ypi\ are almost identical to those of \yps\ (figure~\ref{fig02a}). The only noticeable difference is a positive Weiss temperature, with a low $T$ value $\Theta^{\rm low\,T} = 0.31$\,K indicating weak, but dominantly antiferromagnetic (AFM) exchange. Since the similarity is clearly visible in the data above 1.75\,K and is supported by specific heat data down to lowest temperatures, and because of limited measuring time in the $^3$He equipment, susceptibility and magnetization were not investigated below 1.75\,K.\\
%
%
While the tiny Weiss temperatures point to the weakness of the exchange between 4$f$ moments, the small values might also be the result of competing FM and AFM interactions. This can be discriminated by studying the magnetic specific heat. Within a Heisenberg model for local moment systems, the leading term in the high temperature series expansion for the specific heat is proportional to the sum of the square of all exchange interactions. Therefore, AFM and FM terms do not cancel out as for the Weiss temperature, instead they are additive. Thus the specific heat at higher temperatures and the $T$ dependence of the entropy are more appropriate for the determination of the overall strength of exchange interactions. We have measured the specific heat $C(T)$ of both compounds down to $60\,$mK. Figure~\ref{fig03}({\bf{a}}) shows a log($C/T$) versus log$T$ plot of the measured data in the temperature range between 60\,mK and 4\,K and in a magnetic field of 0\,T and 4\,T. The results are quite similar for both compounds: in zero field $C/T$ shows a clear minimum at around 3.3\,K (\yps) or 2.5\,K (\ypi) from which on it increases continuously with decreasing $T$, following roughly a power law $C/T \propto T^{-2}$ over one decade in temperature. This increase ends in a broaden anomaly at $T_0 = 250\,$mK and $T_0 = 180$\,mK for \yps\ and \ypi, respectively. With further decreasing temperature $C/T$ stays approximately constant before increasing again at lowest $T$. The large $T$ range in which $C/T$ increases with approximately $T^{-2}$ indicates the presence of strong fluctuations. However, this increase is faster than expected for Kondo interaction, since calculations for the single ion model indicate the $T$ dependence of $C(T)$ weaker than $1/T$. Thus attempts to fit the $C(T)$ data with prediction for the single Kondo ion model \cite{Desg82} were not successful. The absence of a significant Kondo interaction is also confirmed by the $T$ dependence of the resistivity (see below). Therefore, this increase of $C/T$ with decreasing $T$ has to be attributed to the emergence of intersite correlations. Accordingly, the broad anomalies at $T_0$ very likely reflect short range order, long range order being prevented by the strong fluctuations. The increase at lowest $T$ can be attributed to nuclear contributions (see below).\\
A magnetic field of just 4\,T has a dramatic effect on the specific heat of both compounds. The whole entropy connected with the continuous increase in $C/T$ and the broad maxima is transferred to a Schottky anomaly with a maximum (in $C/T$) at about 4.5\,K and 2.5\,K in \yps\ and \ypi, respectively. This confirms the intersite exchange to be so weak that a field of 4\,T is completely sufficient to induce a fully polarized state resulting in a corresponding Zeeman splitting of the doublet ground state in agreement to the observed $M(H)$ curve.. As shown later, the involved magnetic entropy is close to $R$\,ln2, indicating that the CEF ground state is a doublet in both compounds \cite{YbPt2InGSdoublet}. The size of the splitting deduced from the temperature of the maximum in $C(T)/T$ roughly corresponds to the value expected from the observed magnetization data. Below 400\,mK the specific heat at 4\,T increases strongly, showing a well-defined $C/T \propto T^{-3}$ power law. Thus, this increase can safely be attributed to nuclear contribution. Calculations (not included) show that this increase can completely be accounted for by the contributions of the two Yb isotopes with a nuclear magnetic moment. At lowest $T$, the data for 4\,T merge with the data for $\mu_{0}H = 0$, indicating that nuclear contributions are similar for $\mu_{0}H =0$ and 4\,T, as one expects for the ground state in a system of localized Kramers ions.\\
Thus both susceptibility $\chi^{-1}(T)$ and specific heat $C(T)/T$ data indicate well-localized Yb$^{3+}$ moments, a doublet CEF ground state, the absence of Kondo interactions, tiny intersite exchange, which further on seems to be frustrated resulting in short range order at about 200\,mK but with intersite correlations surviving up to 2\,K. One can devise several possible origins for the observed strong fluctuations. Yb atoms form a hexagonal lattice in \yps\ and a fcc lattice in \ypi. Both lattice types are prone to geometrical frustration. Furthermore, frustration may also arise from competition between FM and AFM interactions. A further possibility, which is especially relevant for the Heusler structure, is disorder due to partial substitution of Pt by Sn/In and vice versa. Notably strong fluctuations far above a broad magnetic transition at comparatively low temperatures have already been reported and discussed for the Heusler series REPd$_2$Sn and especially for YbPd$_2$Sn \cite{Stan87,Giud04}. On the other hand, in the present compounds some ``intrinsic'' magnetic disorder arise because the hyperfine coupling is not significantly smaller than the weak intersite exchange interactions. Because natural Yb consists of two isotopes with different nuclear moments, Yb$^{171}$ ($I = 1/2$, 14\,\% abundance), Yb$^{173}$ ($I = 5/2$, 16\,\% abundance), as well as 5 isotopes with $I = 0$, hyperfine coupling, once it becomes relevant, will result in three different magnetic Yb species statistically distributed on the single crystallographic Yb site. Our calculations indicate that the overall splitting due to hyperfine coupling can be of the order of 300\,mK for Yb$^{171}$ and 500\,mK for Yb$^{173}$, thus larger than the observed $T_0$.\\
%
%
In order to gain more insight into the strength of exchange interactions, we determined the $T$ dependence of the 4$f$ entropy $S^{4f}(T)$ for both compounds. For this purpose the nuclear contribution was estimated from the experimental data in 4\,T (see above) while the phononic contribution was estimated from the homologue Lu-compounds. Both were subtracted from the data and then $S^{4f}(T)$ was calculated by integrating $C^{4f}(T)/T$. We first note that in zero field $S^{4f}$ saturates above 1\,K at a value of about 5\,J/K\,mol, thus close to the $R$\,ln2 expected for a doublet. This confirms our interpretation of the $C/T$ data and indicates that our estimates of the nuclear parts are reasonable. A first estimation of the strength of exchange interactions can be obtained by taking twice the temperature at which $S^{4f}$ reaches $1/2\,R$\,ln2, which gives an overall scale of about 0.8\,K for both compounds. This value is significantly larger than expected from the Weiss temperatures. As noted above, this discrepancy suggests competition between FM and AFM exchange. An idea about the strength of competing interaction can be obtained from a slightly more involved analysis assuming a Heisenberg model with exchange between nearest neighbors (NN) and next nearest neighbors (NNN). For such a model, the leading term in the $1/T$ series expansion of the specific heat is given by $C = R \cdot 3/32 \cdot 1/T^2 \cdot \sum J_i^2 = \alpha / T^2$, while the Weiss temperature is given by $\Theta = 1/4 \cdot \sum J_i$, where $J_i$ is the exchange strength (in Kelvin) to the neighbor i \cite{John00}. From the evolution of $C/T$ in the temperature range 1\,K $< T <$ 2\,K one can estimate $\alpha$ to be of the order of 1.2\,JK/mol in both compounds. In the hexagonal structure of \yps\ the Yb atom has 6 NN within the hexagonal plane and 6 NNN in the planes above and below. In this compound the negative sign of $\Theta^{\rm low\,T} = -0.08$\,K as well as its small absolute value compared to $\alpha$ implies one of the exchange to be negative, i.e. ferromagnetic, and the other to be positive, i.e. antiferromagnetic. Using the expressions given above, one can easily deduce \cite{standardmath} from the values of $\alpha$ and $\Theta^{\rm low\,T}$ the magnitude of these exchanges: we obtained $J_{FM} = -0.33$\,K and $J_{AFM} = 0.27$\,K. In the cubic Heusler structure \cite{YbPt2Inignore} of \ypi\ the Yb atom has 12 NN and 6 NNN, and we define the respective exchanges as $J_1$ and $J_2$. Because the number of NN and NNN differs, we obtain two solutions from $\alpha = 1.2$\,JK/mol and $\Theta = 0.31$\,K: $J_1 = 0.27$\,K and $J_2 = -0.33$\,K, or $J_1 = -0.13$\,K and $J_2 = 0.47$\,K \cite{standardmath}.\\
All these results point to unusual weak exchange interactions for a dense intermetallic magnetic rare earth compound. One criterion is obviously the small de Gennes factor of Yb, which is e.g. a factor of 50 smaller than that of Gd. For heavy rare earth ($\geq$ Gd) magnetic ordering temperatures roughly scale with the de Gennes factor resulting in comparatively small ordering temperatures in Yb compounds. But even within Yb based intermetallic systems, the strength of the exchanges observed here are very weak. The origin for this weak exchange is still an open question. It is unlikely to be related to specific structural properties, since we observed a similar effect in two completely different structure types. Likely it is more related to the combination of a heavy 5$d$ and a heavy 5$p$ ligands. This very tiny exchange opens a way for a possible application: part ({\bf{b}}) of figure~\ref{fig03} shows a huge shift of the entropy upon applying a comparatively small field of 4\,T. This indicates a huge magnetocaloric effect and suggest these materials to be interesting candidates for adiabatic cooling.
\subsection{Structural transition}
%
%
\begin{figure}[tb]
	\flushright
  \includegraphics[width=0.85\textwidth]{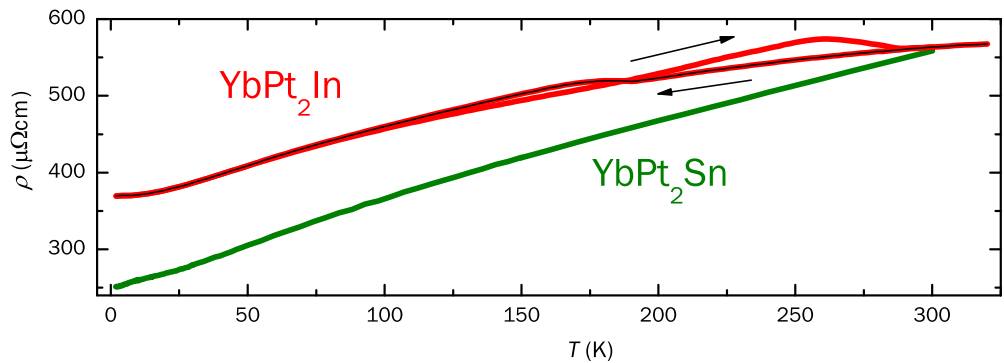}
  \caption{Temperature dependence of the resistivity $\rho$ of \yps\ and \ypi. The positive slope proves a metallic character in both compounds. The anomaly at about 190\,K (decreasing $T$) and 290\,K (increasing $T$) in \ypi\ evidences a first order phase transition, likely of charge density wave type (see text).
  }
 \label{fig04}
\end{figure}
\begin{figure}[tb]
	\flushright
  \includegraphics[width=0.85\textwidth]{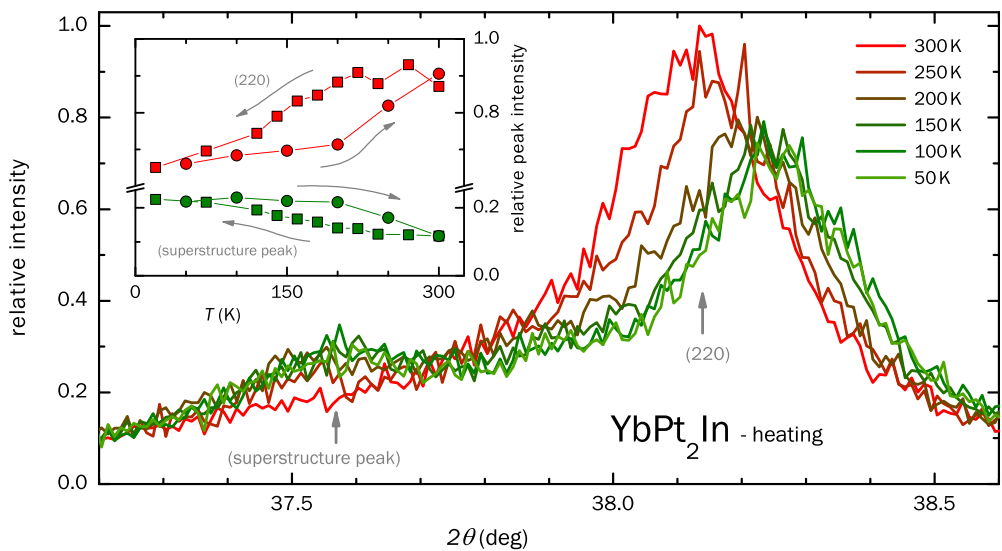}
  \caption{Evolution of the (220) Bragg peak with increasing temperature in \ypi. Above 200\,K the weak superstructure peak at 37.55\,deg starts to disappear proving a structural phase transition. The shift of the central (220) peak to lower angle is due to the normal thermal expansion. The data suggest a further satellite peak near 37.95\,deg.
  }
 \label{fig05}
\end{figure}

%
The temperature dependence of the resistivity $\rho(T)$ shown in figure~\ref{fig04} reveals a continuous increase from lowest to highest temperatures, without strong changes in the slope, except for an anomaly in \ypi\ at around 240\,K upon heating and 170\,K upon cooling. The general behavior of $\rho(T)$ indicates a standard metallic character and the absence of Kondo interaction. On the other hand the anomaly observed in \ypi\ and its large hysteresis suggests the presence of a first order phase transition. This anomaly was reproduced in different samples, although the transition temperature $T_{\rm trans}$ (midpoint) varies from sample to sample from about 240\,K to more than 400\, K. We suspect this sample dependence to be related to slightly different Yb contents since its control is difficult in the arc-melting process. In figure~\ref{fig04} we show the result for the sample with the sharpest kink. Furthermore, its $T_{\rm trans}$ was below 300\,K and thus easily accessible. The upturn in $\rho(T)$ at $T_{\rm trans}$ points to the opening of a gap on part of the Fermi surface and is reminiscent of spin or charge density wave (SDW or CDW) transitions. Since a SDW transition at around 250\,K is very unlikely in view of the weakness of magnetic interactions, this anomaly is an indication for the occurrence of a CDW in \ypi.\\
Therefore, we performed $T$ dependent powder X-ray scattering in order to obtain direct evidence for a structural transition. For these measurements part of the sample used for the $\rho(T)$ curve shown in figure~\ref{fig04} was powdered to a particle size of less than 20\,\textmu m. The diffraction patterns did not reveal any significant splitting of the main peaks of the Heusler phase, but the appearance of several weak additional superstructure peaks below 300\,K. This is demonstrated in figure~\ref{fig05} which shows the strong (220) peak of the Heusler structure and one of the strongest additional peak (superstructure peak) which emerges at a smaller scattering angle for $T < 300$\,K. The evolution of the intensity of both the additional and the main peak reveal a hysteresis as was observed in $\rho(T)$. The appearance of additional peaks at a similar temperature as the anomaly in $\rho(T)$ and with a similar hysteresis confirms the presence of a structural phase transition in \ypi.\\
%
%
For a more precise analysis the diffraction patterns shown in figure~\ref{fig05} and additional ones at further temperatures were fitted with a sum of three Lorentzians: one with all parameter free for the strong (220) peak, one with a fixed line width for the additional low angle superstructure peak, and a third one with all parameter fixed in order to account for a suspected small satellite peak near 37.95\,deg. The position of the (220) peak shifts continuously to higher angle with decreasing $T$, but analysis of the whole diffraction pattern proves that this shift is merely due to the normal lattice contraction upon cooling since all strong peaks shift correspondingly. Furthermore, we could not resolve any significant increase in the width of this peak, indicating that any splitting is very small. In contrast its intensity decreases significantly with $T$, in contradiction to the expected increase due to the increasing Debye Waller factor. Part of this intensity is transferred to the superstructure peak. The finite intensity of this peak at 300\,K is likely related to the problem of determining the correct background, but might also be due to some part of the sample being still in the low $T$ phase at 300\,K. Interestingly, the position of this peak does not shift with $T$, implying that the effect due to the contraction of the lattice is compensated by a change in the structural deformation of the low $T$ phase. The absence of a visible splitting of the main peaks and the appearance of several weak additional peaks indicate that the low $T$ structure is mainly a modulation of the cubic high $T$ structure resulting in a multifold extended cell, as was already suggested for UPd$_2$In \cite{Taka89}. This supports the CDW character of the transition. Because of the small number of superstructure peaks visible in the powder data and their weak intensity, the low $T$ structure could yet not be determined. This requires studies on single crystals, which have now been started. Preliminary studies on LuPt$_2$In also indicate the occurrence of a CDW transition at about 480\,K.\\
%
\begin{figure}[tb]
	\flushright
  \includegraphics[width=0.85\textwidth]{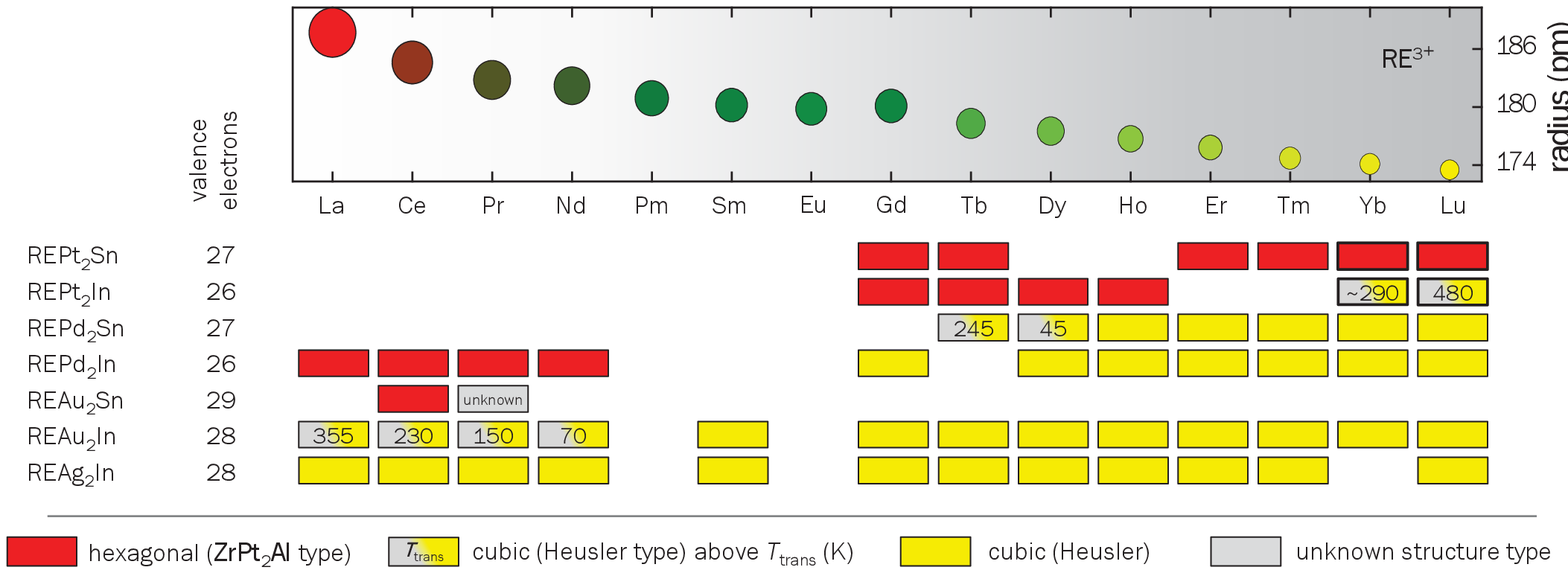}
  \caption{Evolution of the structure in a series of RET$_2$X compounds as a function of rare earth size (References: REPt$_2$Sn~\cite{Dwig87}, REPt$_2$In~\cite{Dwig87,Heyi09},REPd$_2$Sn~\cite{Umar85}, REPd$_2$In~\cite{Bian95,Dwig87a,Baba95}, REAu$_2$Sn~\cite{Mazz09}, REAu$_2$In~\cite{Besn86}, REAg$_2$In~\cite{Gale84}). The four bold framed systems were investigated in the present work. At empty places the compounds have yet not been reported. This overview evidence a systematic change from the hexagonal ZrPt$_2$Al to the cubic Heusler structure type upon decreasing the rare earth size. Just beyond the change the cubic Heusler compounds present a transition to a low $T$ structure. Substituting Pd for Pt strongly favors the Heusler phase, while substituting In for Sn has a much weaker effect. This as well as the comparison with the REAu$_2$In series demonstrates that the valence electron count (VE) is not a relevant parameter for these structural transitions.
  }
 \label{fig06}
\end{figure}
%
Low $T$ structural transitions have already been reported in some RET$_2$X compounds with Heusler structure. Thus, in the series REAu$_2$In a transition has been found for RE~=~La~-~Nd, with $T_{\rm trans}$ decreasing with increasing size of the RE atom \cite{Besn86}. Accordingly all members of this series with RE heavier than Nd stay in the cubic structure down to lowest $T$. Our present observations suggest a systematic trend for the occurrence of this transition. In order to disclose underlying relations we show in figure~\ref{fig06} the evolution of the structure across RET$_2$X series with either ZrPt$_2$Al or Heusler structure types \cite{Dwig87,Umar85,Bian95,Dwig87a,Baba95,Mazz09,Besn86,Gale84}. In horizontal direction the table is arranged according to decreasing size of RE, while in the vertical direction the series are arranged according to increasing stability of the cubic structure. This table evidences several trends: the transition from the hexagonal ZrPt$_2$Al to the Cubic Heusler structure type is promoted by:
	\begin{itemize}
     \item a decrease of the size of RE
     \item replacement of a 5$d$ T metal by a 4$d$ T metal
     \item replacement of a T metal from the Ni column by one of Cu column
     \item replacement of Sn by In
  \end{itemize}
On the other hand, the total number of valence electrons seems to be irrelevant. The table suggests a further correlation: in the compounds with a cubic Heusler structure at high $T$, the structural transition at low $T$ is only observed if the compound is close to the border to the hexagonal structure. Thus, the occurrence of the structural low $T$ transition seems to be related to an incipient instability of the cubic Heusler structure towards a completely different structure type. However the low $T$ structure seems to be quite different for the different series or even within a series. Thus, DyPd$_2$Sn and the REAu$_2$In (RE~=~La~-~Nd) present a strong and well-defined splitting of the cubic peaks indicating a cubic to tetragonal transition, while TbPd$_2$Sn and UPd$_2$In present a splitting in multiple peaks indicating a complex low $T$ structure \cite{Umar85,Besn86,Taka89}. Unfortunately, the low $T$ structure has yet not be determined in any of these compounds.
\section{Conlusion}
We have synthesized the two new compounds \yps\ and \ypi\ and found that they crystallizes in very different structure types, in the hexagonal ZrPt$_2$Al and in the cubic Heusler ones, respectively. Despite different structures the two compounds present very similar magnetic properties indicating a stable trivalent Yb$^{3+}$ experiencing rather weak exchange interactions with a strength well below 1\,K. Specific heat data evidence short range ordering at about 200\,mK, but with strong intersite fluctuations up to 2\,K. The reduced stability of the ordered state is likely due to both geometrical frustration and competing FM and AFM interactions, as indicated by a combined analysis of specific heat and susceptibility results. In Yb-based systems magnetic exchange is expected to be weak because of the small de Gennes factor, but in the present compounds the strength of magnetic interactions is particularly weak compared to other intermetallic Yb-based compounds. This might be related to the ligands being heavy 5$d$ and heavy 5$p$ elements. The weakness of the exchange interactions results in a large magnetocaloric effect, which make these compounds attractive candidates for magnetic cooling.\\
Furthermore, we discovered in \ypi\ a first order phase transition at about 290\,K, to a yet undetermined low $T$ structure which is likely a complex superstructure resulting from a multifold extended cell. At the transition we observe a clear increase in $\rho(T)$ with decreasing $T$, indicating partial gapping of the Fermi surface. This suggests this transition to be a charge density wave. Similar transitions have previously been reported for a few RET$_2$X compounds crystallizing in the cubic Heusler structure. By analyzing the evolution of the structure in several RET$_2$X series of compounds we could relate the occurrence of this low $T$ structural transition to the change of the high $T$ structure from the hexagonal to the cubic one, and disclose the parameters which govern this change in the high $T$ structure.
\ack
This work was financially supported by the Max-Planck-POSTECH Center for complex Phase Materials KR2011-0031558 and by the German Research Foundation (DFG), grant GE602/2-1. The work in Frankfurt was supported through the SFB/TR49. Additional DFG support within the Research Training Group ``Itinerant magnetism and superconductivity in intermetallic compounds'' (DFG-GRK1621) is acknowledged.
\section*{References}
\bibliography{YbPt2X_arXiv}

\begin{thebibliography}{10}
\bibitem{YbPt2InGSdoublet}
In cubic symmetry the $J = 7/2$ multiplet of Yb$^{3+}$ splits into two doublets
  and one quartet, thus a quartet CEF ground state is possible in the cubic
  Heusler structure. However our results indicate a doublet in YbPt$_2$In, in
  accordance with reports on the homologue compounds YbPd$_2$Sn and YbPd$_2$In.
\bibitem{standardmath}
Using standard mathematical operation, we got the following relation between
  the strength of the exchange interactions, the coefficient $\alpha$ of the
  $T^{-2}$ term in $C(T)$, and $\Theta$. We first define the quantities
  $\mathtt{A} = 32\alpha / 18 R$ ($R$ = gas constant) and $\mathtt{B} =
  2\Theta/3$. Then for YbPt$_2$Sn with 6 NN and 6 NNN we obtained $J_{AFM} =
  (\mathtt{B} + \mathtt{C})/2$ and $J_{FM} = (\mathtt{B} - \mathtt{C})/2$ where
  $\mathtt{C} = \sqrt{2\mathtt{A} - \mathtt{B}^2}$. For YbPt$_2$In with 12 NN
  ($J_1$) and 6 NNN ($J_2$), there are two solutions, one with $J_1 =
  (\mathtt{B} + \mathtt{C}) / 3$ and $J_2 = (\mathtt{B} - 2\mathtt{C}) / 3$,
  and one with $J_1 = (\mathtt{B} - \mathtt{C}) / 3$ and $J_2 = (\mathtt{B} +
  2\mathtt{C}) / 3$, where $\mathtt{C} = \sqrt{3\mathtt{A}/2 -
  \mathtt{B}^2/2}$.
\bibitem{YbPt2Inignore}
Here we ignore the structural transition observed at around 300\,K, since the
  related distortions are likely rather small and thus the effect on the
  exchange rather weak.
\end{thebibliography}
\bibliographystyle{unsrt}
\end{document}